\documentclass[12pt]{iopart}
\usepackage{iopams}
\eqnobysec
 
\begin{document}
\jl{1}

\title[Reduced phase space]{Reduced phase space:  
quotienting procedure for gauge theories}

\author{J M Pons\dag
\footnote[2]{E-mail address:  pons@ecm.ub.es},
D C Salisbury\S
\footnote[4]{E-mail address:  dsalisbury@austinc.edu}
and L C Shepley\P
\footnote[6]{E-mail address:  larry@helmholtz.ph.utexas.edu}}

\address{\dag\ Departament d'Estructura i Constituents de la 
Mat\`eria, Universitat de Barcelona, and Institut de F\'\i{}sica 
d'Altes Energies, Diagonal 647, E-08028 Barcelona, Catalonia, Spain}

\address{\S\ Department of Physics, Austin College, Sherman, Texas 
75090-4440, USA}

\address{\P\ Center for Relativity, Physics Department, The University 
of Texas, Austin, Texas 78712-1081, USA}

\begin{abstract}

We present a reduction procedure for gauge theories based on 
quotienting out the kernel of the presymplectic form in 
configuration-velocity space.  Local expressions for a basis of this 
kernel are obtained using phase space procedures; the obstructions to 
the formulation of the dynamics in the reduced phase space are 
identified and circumvented.  We show that this reduction procedure is 
equivalent to the standard Dirac method as long as the Dirac 
conjecture holds: that the Dirac Hamiltonian, containing the primary 
first class constraints, with their Lagrange multipliers, can be 
enlarged to an extended Dirac Hamiltonian which includes all first 
class constraints without any change of the dynamics.  The quotienting 
procedure is always equivalent to the extended Dirac theory, even when 
it differs from the standard Dirac theory.  The differences occur when 
there are ineffective constraints, and in these situations we conclude 
that the standard Dirac method is preferable --- at least for 
classical theories.  An example is given to illustrate these features, 
as well as the possibility of having phase space formulations with an 
odd number of physical degrees of freedom.

\end{abstract}


\vskip 12pt \noindent To appear in \JPA {\bf 32} (1999)

\maketitle


\section{Introduction}
\label{sec.introduction}

The dynamics of gauge theories is a very wide area of research because 
many fundamental physical theories are gauge theories.  The basic 
ingredients are the variational principle, which derives the dynamics 
out of variations of an action functional, and the gauge principle, 
which is the driving principle for determining interactions based on a 
Lie group of internal symmetries.  The gauge freedom exhibited by the 
complete theory becomes a redundancy in the physical description.  The 
classical treatment of these systems was pioneered by Dirac (1950, 
1964), Bergmann (1949), and Bergmann and Goldberg (1955).  Dirac's 
classical treatment in phase space (the cotangent bundle for 
configuration space) has been shown (Gotay and Nester 1979, 1980, 
Batlle \etal 1986) to be equivalent to the Lagrangian formulation in 
configuration-velocity space (the tangent bundle).  One ends up with a 
constrained dynamics with some gauge degrees of freedom.  One may 
choose, as is customary in many approaches (Pons and Shepley 1995) 
to introduce new constraints in the formalism to eliminate these 
unwanted --- spurious --- degrees of freedom.  This is the gauge 
fixing procedure.

There are approaches other than gauge fixing.  For instance, the 
method of Faddeev and Jackiw (1993)and Jackiw (1995) is to 
attempt to reduce the system to its physical degrees of 
freedom by a process of directly substituting the constraints 
into the canonical Lagrangian.  It has been proved 
(Garc\'\i{}a and Pons 1997) that, as long as ineffective constraints 
--- functions that vanish in the constraint surface and whose 
differentials also vanish there --- are not present, the 
Faddeev-Jackiw method is equivalent to Dirac's.

A reduction procedure (Abraham and Marsden 1978, Sniatycki 1974, 
Lee and Wald 1990) which seems to be particularly appealing from 
a geometric point of view consists in quotienting out the kernel of 
the presymplectic form in configuration-velocity space in order to get 
a reduced space, the putative physical space, with a deterministic 
dynamics in it, that is, without gauge degrees of freedom.  One must 
be careful that these techniques do not change the physics, for 
example by dropping degrees of freedom, and that they are applicable 
in all situations of physical interest.  For example, we know of no 
treatment of this technique which applies to the important case when 
there are secondary constraints --- one purpose of this paper is to 
provide this treatment.

In this paper we present a complete reduction method based on 
quotienting out the kernel of the presymplectic form.  We establish a 
systematic algorithm and prove its equivalence with Dirac's method, 
but only so long as ineffective constraints do not appear.  Our 
procedure turns out to be equivalent to Dirac's extended method, which 
enlarges the Hamiltonian by including all first class constraints.  It 
differs from the ordinary Dirac method (supplemented by gauge fixing) 
when ineffective constraints occur.  Since the ordinary Dirac method 
is equivalent to the Lagrangian formalism, it is to be preferred in 
classical models.

We will consider Lagrangians with gauge freedom.  Thus they must be 
singular: The Hessian matrix of the Lagrangian, consisting of its 
second partial derivatives with respect to the velocities, is 
singular, or equivalently, the Legendre transformation from 
configuration-velocity space to phase space is not locally invertible.  
Singular also means that the pullback under this map of the canonical 
form $\bomega$ from phase space to configuration-velocity space is 
singular.

In order to proceed, we first compute, in section \ref{sec.kernel}, in 
a local coordinate system, a basis for the kernel of the presymplectic 
form.  Our results will be in general local; global results could be 
obtained by assuming the Lagrangian to be almost regular (Gotay and 
Nester 1980).  In section \ref{sec.obstructions}, we will single out 
the possible problems in formulating the dynamics in the reduced space 
obtained by quotienting out this kernel.  In section 
\ref{sec.physical} we will solve these problems and will compare our 
results with the classical Dirac method.  It proves helpful to work in 
phase space here, and we end up with a reduced phase space complete 
with a well-defined symplectic form.  In section \ref{sec.example} we 
illustrate our method with a simple example (which contains 
ineffective constraints).  We draw our conclusions in section 
\ref{sec.conclusions}.


\section{The kernel of the presymplectic form}
\label{sec.kernel}

We start with a singular Lagrangian $L(q^i,\dot q^i)$ 
$(i=1,\cdots,N)$.  The functions 
\[\hat p_i(q,\dot q):=\partial L/\partial\dot q^i
\]
are used to define the Hessian $W_{ij}=\partial\hat p_i/\partial\dot 
q^j$, a singular matrix that we assume has a constant rank $N-P$.  The 
Legendre map ${\cal F}\!L$ from configuration-velocity space (the 
tangent bundle) $TQ$ to phase space $T^*\!Q$, defined by $p_i=\hat 
p_i(q,\dot q)$, defines a constraint surface of dimension $2N-P$.

The initial formulation of the dynamics in $TQ$ uses the Lagrangian 
energy 
\[  E_{\rm L} := \hat p_i\dot q^i-L
\]
and {\bf{}X}, the dynamical vector field on $TQ$,
\begin{equation}
    i_\mathbf{X} \bomega_{\rm L} = \mathbf{d}(E_{\rm L}) \ ,
    \label{x}
\end{equation}
where 
\[  \bomega_{\rm L} := \mathbf{d}q^s \wedge \mathbf{d}\hat p_{s}
\]
is the pullback under the Legendre map of the canonical form 
$\bomega=\mathbf{d}q^{s}\wedge\mathbf{d}p_{s}$ in phase space.  
$\bomega_{\rm L}$ is a degenerate, closed two-form, which is called 
the presymplectic form on $TQ$.  In fact there is an infinite number 
of solutions for equation \eref{x} if the theory has gauge freedom, 
but they do not necessarily exist everywhere (if there are Lagrangian 
constraints).  {\bf{}X} must obey the second order condition for a 
function:
\[ \mathbf{X} q^{i} = \dot q^{i}\ \Longleftrightarrow \mathbf{X} 
    = \dot q^{s}{\partial\over\partial q^{s}} + A^{s}(q,\dot q)
        {\partial\over\partial \dot q^{s}}\ ,
\]
where $A^{s}$ is partially determined by equation \eref{x}.

At first sight, the kernel of $\bomega_{\rm L}$ describes, in 
principle, the arbitrariness in the solutions {\bf{}X} of equation 
\eref{x}.  Therefore it is tempting to think that in order to 
construct a physical phase space, we must just quotient out this 
kernel.  The complete implementation of this procedure, which we
are presenting in this paper is, first, far from obvious, and 
second, as we will see, fraught with danger.

Let us first determine a basis for 
\[  {\cal K} := {\rm Ker}(\bomega_{\rm L})
\] 
in local coordinates.  We look for vectors {\bf{}Y} satisfying
\begin{equation}
    i_\mathbf{Y} \bomega_{\rm L} = 0 \ .
    \label{kern}
\end{equation}
With the notation
\[  \mathbf{Y} = \epsilon^k {\partial\over\partial q^k} + 
        \beta^k {\partial\over\partial\dot q^k}  \ ,  
\]
equation \eref{kern} implies
\numparts\label{ab}
\begin{eqnarray}
    \epsilon^i W_{ij} & = & 0\ , 
    \label{a}  \\
    \epsilon^i A_{ij} - \beta^i W_{ij} & = & 0\ ,
    \label{b}  
\end{eqnarray}
\endnumparts 
where
\[  A_{ij} := {\partial\hat p_i\over\partial q^j} 
    - {\partial \hat p_j\over\partial q^i}\ . 
\]

Since {\bf W} is singular (this causes the degeneracy of 
$\bomega_{\rm L}$), it possesses null vectors.  It is very 
advantageous to this end to use information from phase space.  It is 
convenient to use a basis for these null vectors, $\gamma^i_\mu$, 
$(\mu=1,\dots,P)$, which is provided from the knowledge of the $P$ 
primary Hamiltonian constraints of the theory, $\phi^{(1)}_\mu$.  
Actually, one can take (Batlle \etal 1986),
\begin{equation}
    \gamma^i_\mu\ 
    = {\cal F}\!L^*
        \left({\partial\phi^{(1)}_\mu\over\partial p_i}\right)
    = {\partial\phi^{(1)}_\mu\over\partial p_i}(q,\hat p) \ ,
    \label{gamma} 
\end{equation}
where ${\cal F}\!L^*$ stands for the pullback of the Legendre map 
${\cal F}\!L: T Q \longrightarrow T^*\!Q$.  According to equation 
\eref{a}, $\epsilon^i$ will be a combination of these null vectors: 
$\epsilon^i = \lambda^\mu \gamma^i_\mu$.  Notice that we presume that 
these primary constraints are chosen to be effective.

To have a solution for $\beta^i$ we need, after contraction of 
equation \eref{b} with the null vectors $\gamma^j_\nu$, 
\[  0 = \lambda^\mu \gamma^i_\mu A_{ij} \gamma^j_\nu 
    = \lambda^\mu {\cal F}\!L^*
        \left({\partial\phi^{(1)}_\mu\over\partial p_i}\right) 
        \left({\partial\hat p_i\over\partial q^j}
       - {\partial\hat p_j\over\partial q^i}\right)
            {\cal F}\!L^*
        \left({\partial\phi^{(1)}_\nu\over\partial p_j}\right)\ ,
 \]
which is to be understood as an equation for the $\lambda^\mu$s. 
We then use the identity
\begin{equation}
	{\cal F}\!L^*
	        \left({\partial\phi^{(1)}_\mu\over\partial p_j}\right) 
	        {\partial{\hat p}_j\over \partial q^i}
	    + {\cal F}\!L^*
	        \left({\partial\phi^{(1)}_\mu\over\partial q^i}\right)
	    = 0\ ,
	\label{idenx}
\end{equation}
which stems from the fact that $\phi^{(1)}_\mu (q,\hat p)$ vanishes 
identically; we get
\begin{eqnarray}
    0 & = &
    \lambda^\mu {\cal F}\!L^*
        \left({\partial\phi^{(1)}_\mu\over\partial p_i} 
        {\partial\phi^{(1)}_\nu\over\partial q^i}
    - {\partial\phi^{(1)}_\mu\over\partial q^j}
        {\partial\phi^{(1)}_\nu\over\partial p_j}\right)
    \nonumber \\
    & = & \lambda^\mu {\cal F}\!L^*
        \left(\{\phi^{(1)}_\nu,\phi^{(1)}_\mu\}\right) \ . 
    \label{fc} 
\end{eqnarray}
Condition \eref{fc} means that the combination 
$\lambda^\mu\phi^{(1)}_\mu$ must be first class.  Let us split the 
primary constraints $\phi^{(1)}_\mu$ between first class 
$\phi^{(1)}_{\mu_1}$ and second class $\phi^{(1)}_{\mu'_1}$ at the 
primary level, and we presume that second class constraints are second 
class everywhere on the constraint surface (more constraints may 
become second class if we include secondary, tertiary, etc, 
constraints).  They satisfy
\begin{equation}
    \{ \phi^{(1)}_{\mu_1},\,\phi^{(1)}_{\mu} \} = pc\ , \quad 
    \det |\{\phi^{(1)}_{\mu'_1},\phi^{(1)}_{\nu'_1} \}| \not= 0\ , 
    \label{fc-sc} 
\end{equation}
where $pc$ stands for a generic linear combination of primary 
constraints.  Equations \eref{fc} simply enforce 
\[  \lambda^{\mu'_1} = 0 \ .  
\]
Consequently a basis for the $\epsilon^i$ will be spanned by 
$\gamma_{\mu_1}$, so that 
\[  \epsilon^i=\lambda^{\mu_1}\gamma^i_{\mu_1}
\]
for $\lambda^{\mu_1}$ arbitrary.  Once $\epsilon^i$ is given, 
solutions for $\beta^i$ will then be of the form
\[ \beta^i 
    = \lambda^{\mu_1}\beta^i_{\mu_1} + \eta^\mu \gamma^i_\mu\ ,
\] 
where the $\eta^\mu$ are arbitrary functions on $TQ$.  We will now 
determine the $\beta^j_{\mu_1}$.

To compute $\beta^j_{\mu_1}$ it is again very convenient to use 
Hamiltonian tools.  Consider any canonical Hamiltonian $H_{\rm c}$ 
(which is not unique), that is, one satisfying $E_{\rm L} = {\cal 
F}\!L^*(H_{\rm c})$.  Since we know from the classical Dirac analysis 
that the first class primary constraints $\phi^{(1)}_{\mu_1}$ may 
produce secondary constraints,

\[  \phi^{(2)}_{\mu_1}= \{\pi^{(1)}_{\mu_1},H_{\rm c} \}\ ,
\] 
we compute (having in mind equation \eref{b})
\begin{eqnarray}
    \gamma^i_{\mu_1} A_{ij} +
    {\partial\phi^{(2)}_{\mu_1}\over\partial p_i}(q,\hat p) W_{ij} 
    &=& \gamma^i_{\mu_1} A_{ij} 
        + {\partial\phi^{(2)}_{\mu_1}\over\partial p_i}(q,\hat p) 
            {\partial\hat p_i\over\partial \dot q_j}
    \nonumber\\ 
    &=& \gamma^i_{\mu_1} A_{ij}
        + {\partial{\cal F}\!L^*(\phi^{(2)}_{\mu_1})
            \over\partial\dot q_j}
    \nonumber\\ 
    &=& \gamma^i_{\mu_1} A_{ij} 
        + {\partial (K\phi^{(1)}_{\mu_1})\over\partial\dot q_j}\ , 
    \label{nozero} 
\end{eqnarray}
where we have used the operator $K$ 
defined (Batlle \etal 1986, Gr\`acia and Pons 1989) by
\[  K f := \dot q^i {\cal F}\!L^*
            \left({\partial f\over\partial q^i}\right)
        + {\partial L \over \partial q^i} 
            {\cal F}\!L^*
                \left({\partial f\over\partial p_i}\right) \ . 
\]
This operator satisfies (Batlle \etal 1986, Pons 1988) 
\begin{equation} 
    K f = {\cal F}\!L^*
            \left(\{f,H_{\rm c} \}\right) 
        + v^\mu(q,\dot q) {\cal F}\!L^*
            \left(\{f,\phi^{(1)}_\mu \}\right) \ ,
    \label{prop}
\end{equation}
where the functions $v^\mu$ are defined through the identities 
\begin{equation} 
    \dot q^i = {\cal F}\!L^* \left(\{q^i,H_{\rm c}\}\right)
        + v^\mu(q, \dot q) 
            {\cal F}\!L^*\left(\{q^i, \phi_\mu^{(1)} \}\right)\ .
    \label{propv}
\end{equation}
Property \eref{prop} implies, for our first class constraints, 
\[  K \phi^{(1)}_{\mu_1}
    = {\cal F}\!L^*\left(\phi^{(2)}_{\mu_1}\right) \ , 
\] 
which has been used in equation \eref{nozero}.  Let us continue with 
equation \eref{nozero}:
\begin{eqnarray}
    \gamma^i_{\mu_1} A_{ij}  +  
        {\partial (K\phi^{(1)}_{\mu_1})\over\partial\dot q_j} & = &
     - {\cal F}\!L^*
         \left({\partial\phi^{(1)}_{\mu_1}\over\partial q^j}\right) 
       - {\cal F}\!L^* 
         \left({\partial\phi^{(1)}_{\mu_1}\over\partial p_i}\right)
         {\partial\hat p_j\over\partial q^i}
     \nonumber \\
     & & + {\partial\over\partial\dot q^j}
         \left(\dot q^{i}{\cal F}\!L^*
             \left({\partial \phi^{(1)}_{\mu_1}\over
                 \partial q^{i}}\right) 
         +{\partial L \over \partial q^{i}}{\cal F}\!L^*
             \left({\partial \phi^{(1)}_{\mu_1}\over
                 \partial p_{i}}\right)\right)
     \nonumber\\ 
    &= & W_{ij} K {\partial\phi^{(1)}_{\mu_{1}}\over
        \partial p_{i}} \ ,
    \label{ppprop} 
\end{eqnarray}
where we have omitted some obvious steps to produce the final 
result.  We can read off from this computation the solutions for 
equation \eref{b}:
\begin{equation}
    \beta^j_{\mu_1} = K {\partial\phi^{(1)}_{\mu_1}\over\partial p_j} 
        - {\cal F}\!L^*\left({\partial\phi^{(2)}_{\mu_1}
            \over\partial p_j}\right) \ . 
\end{equation} 

Therefore, a basis for $\cal K$ is provided by: 
\numparts\label{nul}
\begin{equation}
    \bGamma_\mu := \gamma^j_\mu 
        {\partial\over\partial{\dot q}^j} 
    \label{nul1} 
\end{equation}
and
\begin{equation}
    \bDelta_{\mu_1} := \gamma^j_{\mu_1} 
        {\partial\over\partial q^j} 
    +\beta^j_{\mu_1} {\partial\over\partial{\dot q}^j} \ .
    \label{nul2}
\end{equation}
\endnumparts Vectors $\bGamma_\mu$ in equation \eref{nul1} form a 
basis for ${\rm Ker}(T{\cal F}\!L)$, where $T{\cal F}\!L$ is the 
tangent map of ${\cal F}\!L$ (also often denoted by ${\cal 
F}\!L_{*}$).  They also span the vertical subspace of $\cal K$: ${\rm 
Ker}(T{\cal F}\!L) = {\rm Ver}({\cal K})$.  This is a well known 
result (Cari\~{n}ena \etal 1988), but as far as we know equations 
(\ref{nul1}, \ref{nul2}) are the first explicit local expression for 
$\cal K$ itself.

All other results (Cari\~{n}ena 1990), obtained on geometrical 
grounds, for $\cal K$ are obvious once the basis for this kernel is 
displayed, as it is in equations (\ref{nul1}, \ref{nul2}).  For 
instance, it is clear that $dim\,{\cal K}\leq2\,dim\,{\rm Ver}({\cal 
K})$.  Also, defining the vertical endomorphism
\[  \mathbf{S} = {\partial\over\partial\dot q^i} 
        \otimes \mathbf{d}q^i\ , 
\]
we have $\mathbf{S}({\cal K)} \subset {\rm Ver}({\cal K})$.  The case 
when
\[  \mathbf{S}({\cal K}) = {\rm Ver}({\cal K})\ ,
\]
corresponds, in the Hamiltonian picture, to the case when all primary 
constraints are first class (indices $\mu$ = indices ${\mu_1}$).  
These are the so-called Type II Lagrangians (Cantrjin \etal 1986).  
$\mathbf{S}({\cal K}) = \emptyset $ corresponds to the case when all 
primary constraints are second class (indices $\mu$ = indices 
${\mu'_1}$).

\Eref{nul1} implies, for any function $f$ on $T^{*}\!Q$,
\begin{equation}
	\bGamma_{\mu}\left({\cal F}\!L^{*}(f)\right) = 0\ .
\end{equation}
The corresponding equation for $\bDelta_{\mu_{1}}$ is:
\begin{equation}
    \bDelta_{\mu_{1}}\left({\cal F}\!L^*(f)\right)
    = {\cal F}\!L^*\left(\{f,\phi^{(1)}_{\mu_1}\}\right) \ .
    \label{delprop}
\end{equation}
Since we will make use of this property below, we now prove this 
result.  The action of $\bDelta_{\mu_{1}}$ is
\begin{eqnarray*}
    \bDelta_{\mu_{1}}\left({\cal F}\!L^*(f)\right)
        & = &\gamma^j_{\mu_1} \left({\cal F}\!L^*
                \left({\partial f\over\partial q^{j}}\right)
          + {\partial\hat p_{i}\over\partial q^{j}}
            {\cal F}\!L^*
               \left({\partial f\over\partial p_{i}}\right)\right) \\
        &  & \quad    +\beta^j_{\mu_1} W_{ji}{\cal F}\!L^*
                \left({\partial f\over\partial p_{i}}\right)\ .
\end{eqnarray*}
We use equations \eref{b}, \eref{gamma}, and \eref{idenx} to get
\begin{eqnarray*}
    \bDelta_{\mu_{1}}\left({\cal F}\!L^*(f)\right)
        & = & {\cal F}\!L^*
            \left({\partial\phi^{(1)}_{\mu_{1}}\over\partial p_{j}}
            {\partial f\over\partial q^{j}}\right)
          - {\cal F}\!L^*
            \left({\partial\phi^{(1)}_{\mu_{1}}\over\partial q^{j}}
            {\partial f\over\partial p_{j}}\right) \\
    & = & {\cal F}\!L^*\left(\{f,\phi^{(1)}_{\mu_{1}}\}\right) \ .
\end{eqnarray*}

The commutation relations (Lie Brackets) for the vectors in equations 
(\ref{nul1}, \ref{nul2}) are readily obtained, and we present these 
new results here for the sake of completeness.  We introduce the 
notation
\begin{eqnarray*}
    \{ \phi_{\mu_1},\phi_{\mu} \} & = & 
        A_{{\mu_1}\mu}^\nu\phi_{\nu} \ , \\
    \{ \phi_{\mu_1},\phi_{\nu_1} \} & = & 
        B_{{\mu_1}{\nu_1}}^{\rho_1}\phi_{\rho_1}
        +{1\over2}B_{{\mu_1}{\nu_1}}^{\rho \sigma}
            \phi_{\rho}\phi_{\sigma}
\end{eqnarray*}
(commutation of first class constraints is also first class).  We 
arrive at
\numparts\label{comm22}
\begin{eqnarray}
    ~[\bGamma_\mu,\bGamma_\nu] & = & 0 \ ,
    \label{comm22a}  \\ 
    ~[\bGamma_\mu,\bDelta_{\mu_1}] & = & 
            {\cal F}\!L^* 
            \left(A_{{\mu_1}\mu}^\nu\right) \bGamma_{\nu} \ ,
        \label{comm22b}  \\
    ~[\bDelta_{\mu_1},\bDelta_{\nu_1}] & = & 
        {\cal F}\!L^* \left(B_{{\nu_1}{\mu_1}}^{\rho_1}\right)
            \bDelta_{\rho_1}
      + v^{\delta'_1}{\cal F}\!L^* 
             \left(B_{{\nu_1}{\mu_1}}^{\rho{\sigma'_1}} 
             \{\phi_{\sigma'_1},\phi_{\delta'_1} \}\right) 
                 \bGamma_\rho \ ,
    \label{comm22c}
\end{eqnarray}
\endnumparts where the $v^{\delta'_1}$ are defined in equation 
\eref{propv}.  Observe that the number of vectors in equations 
(\ref{nul1}, \ref{nul2}) is even because $|\mu'_1| = |\mu| - |\mu_1|$ 
is the number of second class primary constraints (at the primary 
level), which is even.

Because the algebra of $\cal K$ is closed, the action of $\cal K$ on 
$TQ$ is an equivalence relation.  We can form the quotient space 
$TQ/{\cal K}$ and the projection
\[  \pi : TQ \longrightarrow TQ/{\cal K}\ .
\]  
$TQ/{\cal K}$ is endowed with a symplectic form obtained by 
quotienting out the null vectors of $\bomega_{\rm L}$ (that is, 
$\bomega_{\rm L}$ is projectable to $TQ/{\cal K}$).  The space 
$TQ/{\cal K}$ is not necessarily the final physical space, however, 
because we have not yet formulated the dynamics of the system: We 
now turn to the question of the projectability of the Lagrangian 
energy.


\section{Obstructions to the projectability of the Lagrangian energy} 
\label{sec.obstructions}

In order to project the dynamical equation \eref{x} to $TQ/{\cal K}$, 
we need $E_{\rm L}$ to be projectable under $\pi$.  However, in order 
for $E_{\rm L}$ to be projectable we must check whether it is constant 
on the orbits generated by $\cal K$ as defined by the vector fields of 
equations (\ref{nul1}, \ref{nul2}).  Indeed $\bGamma_\mu (E_{\rm 
L})=0$, but from equation \eref{delprop},
\[  \bDelta_{\mu_1} (E_{\rm L}) 
    = - {\cal F}\!L^*\left(\phi^{(2)}_{\mu_1}\right) \ ,
\] 
where 
\[  \phi^{(2)}_{\mu_1}:=\{\phi^{(1)}_{\mu_1},H_{\rm c}\}\ .  
\]
If ${\cal F}\!L^*(\phi^{(2)}_{\mu_1})\neq 0$ for some $\mu_1$, then 
$\phi^{(2)}_{\mu_1}$ is a secondary Hamiltonian constraint.  As a side 
remark, note that in this case ${\cal F}\!L^*(\phi^{(2)}_{\mu_1}$) is 
a primary Lagrangian constraint.  In fact it can be written (Batlle 
\etal 1986) as
\[  {\cal F}\!L^*\left(\phi^{(2)}_{\mu_1}\right) 
    = [L]_i \gamma^i_{\mu_1} \ ,
\]
where $[L]_i$ is the Euler-Lagrange functional derivative of $L$.

We see that there is an obstruction to the projectability of 
$E_{\rm L}$ to $TQ/{\cal K}$ as long as there exist secondary 
Hamiltonian constraints or equivalently if there exist 
Lagrangian constraints.

One way to remove this problem (Ibort and Mar\'{\i}n-Solano 1992), 
Ibort \etal 1993) is to use the coisotropic embedding theorems 
(Gotay 1982), Gotay and Sniatycki 1981) and look for an extension of 
the tangent space possessing a regular Lagrangian that extends the 
singular one and leads to a consistent theory once the extra degrees 
of freedom are removed.  This method is equivalent to Dirac's, but 
only if there are no secondary Hamiltonian constraints.  However, if 
there are, which is precisely our case, the dynamics becomes modified 
and thus changes the original variational principle.  Instead of using 
this technique we will try to preserve the dynamics.


\section{Physical space}
\label{sec.physical}

In the cases where secondary Hamiltonian constraints do exist (for 
instance, Yang-Mills and Einstein-Hilbert theories), we must find an 
alternative reduction of $TQ$ in order to obtain the projectability of 
$E_{\rm L}$.

The initial idea was to quotient out the orbits defined by equations 
(\ref{nul1}, \ref{nul2}).  Since \hbox{$\bGamma_\mu (E_{\rm L})=0$} we 
can at least quotient out the orbits defined by equation \eref{nul1}.  
But this quotient space, ${TQ/{\rm Ker}(T{\cal F}\!L)}$, is already 
familiar to us: It is isomorphic to the surface $M_{1}$ defined by the 
primary constraints in $T^*\!Q$.  In fact, if we define $\pi_1$ as the 
projection
\[  \pi_1: TQ \longrightarrow {TQ/{\rm Ker}(T{\cal F}\!L)}\ ,
\]
we have the decomposition of the Legendre map 
${\cal F}L = i_1 \circ \pi_1$, where 
\[  i_1 : {TQ\over{\rm Ker}(T{\cal F}\!L)} \longrightarrow T^*\!Q\ ,
\]
with 
\[  i_{1}\left({TQ\over{\rm Ker}(T{\cal F}\!L)}\right)=M_{1}\ .
\]

Now we can take advantage of working in $M_1\subset T^*\!Q$.  Let us 
project our original structures on $TQ$ into $M_1$.  Consider the 
vector fields $\bDelta_{\mu_1}$.  \Eref{delprop} tells 
us that the vector fields $\bDelta_{\mu_1}$ are projectable to 
$M_1$ and that their projections are just $\{-,\phi^{(1)}_{\mu_1}\}$.  
In fact these vector fields $\{-,\phi^{(1)}_{\mu_1}\}$ are vector 
fields of $T^*\!Q$, but they are tangent to $M_1$ because 
$\phi^{(1)}_{\mu_1}$ are first class (among the primary constraints 
defining $M_1$).  Incidentally, note that the vector fields 
$\{-,\phi^{(1)}_{\mu'_1}\}$ associated with the second class primary 
constraints in $T^*\!Q$ are not tangent to $M_1$.  

Formulation in $M_1$ of the dynamics corresponding to equation 
\eref{x} uses the pre-symplectic form $\bomega_1$ defined by 
$\bomega_1 = i^*_1 \bomega$, where $\bomega$ is the 
canonical form in phase space, and the Hamiltonian 
$H_1$ defined by $H_1 = i^*_1 H_{\rm c}$, with 
$H_{\rm c}$ such that ${\cal F}\!L^*(H_{\rm c})= E_{\rm L}$.  The 
dynamic equation in $M_1$ will be:
\begin{equation}
    i_{\mathbf{X}_1} \bomega_1 = \mathbf{d}H_1\ .
    \label{x1}
\end{equation}

The null vectors for $\bomega_1$ are $\{-,\phi^{(1)}_{\mu_1}\}$
(more specifically, their restriction to $M_1$).  (This result 
is general:  The kernel of the pullback of the symplectic 
form to a constraint surface in $T^{*}\!Q$ is locally 
spanned by the vectors associated, through the Poisson 
Bracket, with the first class constraints among the constraints which 
define the surface.)  To project the dynamics of equation 
\eref{x1} to the quotient of $M_1$ by the orbits defined by 
$\{-,\phi^{(1)}_{\mu_1}\}$:
\begin{equation}
    {\cal P}_{1} := {M_{1} \over (\{-,\phi^{(1)}_{\mu_{1}}\})} \ ,
    \label{P.1}
\end{equation}
we need the projectability of $H_1$ to this quotient manifold.  To 
check this requirement it is better to work in $T^*\!Q$.  Then 
projectability of $H_1$ to ${\cal P}_{1}$ is equivalent to requiring 
that $\{\phi^{(1)}_{\mu_1},H_{\rm c}\}|{}_{{}_{M_1}}=0$.

Here lies the obstruction we met in the previous section, for it is 
possible that $\{\phi^{(1)}_{\mu_1},H_{\rm c}\}|{}_{{}_{M_1}}\neq 0$ 
for some constraints $\phi^{(1)}_{\mu_1}$.  Let us assume that this is 
the case.  As we did before, we define
\[  \phi^{(2)}_{\mu_1}:=\{\phi^{(1)}_{\mu_1},H_{\rm c}\} \ .
\] 
These constraints may not be independent, some of them may vanish 
on $M_1$, and some previously first class constraints may become 
second class.  Those that do not vanish are secondary constraints and 
allow us to define the new surface $M_2 \subset M_1$ (we define the 
map $i_2: M_2 \longrightarrow M_1$) by $\phi^{(2)}_{\mu_1} = 0$.

We can now form the projection of $H_2:= i_2^*H_1$ to 
$M_2/(\{-,\phi^{(1)}_{\mu_1}\})$, but the projection of $\bomega_2:= 
i_2^*\bomega_1$ can be still degenerate in this quotient space, since 
$\bomega_2$ may have acquired new null vectors (and may have lost some 
of the old ones).  In fact, once all constraints are expressed in 
effective form, ${\rm Ker}(\bomega_2)$ is generated under the Poisson 
Bracket associated with $\bomega$ by the subset of effective 
constraints that are first class with respect to the whole set of 
constraints defining $M_2$.  If there is a piece in this kernel that 
was not present in ${\rm Ker}(\bomega_1)$, then new conditions for the 
projectability of $H_2$ will appear.

The dynamic equation in $M_2$ is
\begin{equation}
    i_{\mathbf{X}_2} \bomega_2 = \mathbf{d}H_2 \ .
    \label{x2}
\end{equation}
It is still convenient to work with structures defined in $T^*\!Q$.  
Suppose that $\phi^{(2)}_{\mu_2}$ is any secondary, first class, 
effective constraint in $M_2$; therefore $\{-,\phi^{(2)}_{\mu_2}\}
\in {\rm Ker}(\bomega_2)$ but $\{-,\phi^{(2)}_{\mu_2}\} 
\notin {\rm Ker}(\bomega_1)$.  The new projectability condition for 
$H_2$ induced by $\phi^{(2)}_{\mu_2}$ is
\[  \{\phi^{(2)}_{\mu_2} ,H_{\rm c}\}|{}_{{}_{M_2}} = 0\ .
\]
This means that we might find new constraints if this 
condition is not satisfied.  A new surface $M_3$ 
will appear, and a new kernel for a new $\bomega_3$ should be 
quotiented out, and so on.  We will not go further because we are just 
reproducing Dirac's algorithm in phase space (Dirac 1950, 1964, 
Batlle \etal 1986, Gotay \etal 1978).  We do have a shift of 
language, however: What in Dirac's standard algorithm is regarded as a 
condition for the Hamiltonian vector field to be tangent to the 
constraint surface is here regarded as a projectability condition for 
the Hamiltonian to a quotient space.

To summarize:  The constraint surface $M_{1}$ is defined by the 
primary constraints $\phi^{(1)}_{\mu}$, a subset of which are the 
first class constraints $\phi^{(1)}_{\mu_{1}}$.  These first class 
constraints are used in the formation of the quotient space
\[  {\cal P}_{1} = {M_{1}\over\{-,\phi^{(1)}_{\mu_{1}}\}} \ .
\]
The projectability condition for $H_{1}$ (the pullback of $H_{\rm c}$ 
to $M_1$) to ${\cal P}_{1}$ may be expressed as the condition 
$\{H_{\rm c},\phi^{(1)}_{\mu_{1}}\}|_{M_{1}}=0.$ If this condition 
holds, we have found the final physical space.  If it doesn't, there 
are new, secondary constraints $\phi^{(2)}_{\mu_{1}}$, and these 
constraints along with the initial set of primary constraints 
$\phi^{(1)}_{\mu}$ are used to define a constraint surface $M_{2}$.  
Among the set of constraints used to define $M_{2}$ are first class 
constraints, including a subset of the first class constraints 
associated with $M_{1}$, which we denote $\phi^{(1)}_{\mu_{2}}$, and a 
subset of the secondary constraints, which we denote 
$\phi^{(2)}_{\mu_{2}}$.  These first class constraints are used in the 
formulation of the quotient space
\[  {\cal P}_{2} := {M_{2}\over(\{-,\phi^{(1)}_{\mu_{2}}\}, 
         \{-,\phi^{(2)}_{\mu_{2}}\})} \ .
\]
Again we must require projectability of the Hamiltonian; eventually, 
the final phase space is
\begin{equation}
    {\cal P}_f := {M_f \over 
    (\{-,\phi^{(1)}_{\mu_{f}}\}, 
     \{-,\phi^{(2)}_{\mu_{f}}\}, 
      \dots, 
     \{-,\phi^{(k)}_{\mu_{f}}\}) } \ ,
    \label{phys.space}
\end{equation}
where $\phi^{(n)}_{\mu_{f}}$ are the final first class $n$-ary 
constraints, all of which are taken in effective form.  ${\cal P}_f$ 
is endowed with a symplectic form which is the projection of the form 
$\bomega_f$ in $M_f$, which is the final constraint surface.  
The dimension of ${\cal P}_f$ is $2N-M-P_f$, 
where $N$ is the dimension of the initial configuration space, $M$ is 
the total number of constraints, and $P_f$ is the number of final 
first class constraints.  Observe that we end up with the standard 
counting of degrees of freedom for constrained dynamical systems: 
First class constraints eliminate two degrees of freedom each, whereas 
second class constraints eliminate only one each.  The final result is 
an even number because the number of second class constraints is even.

In order to use the technique we've presented, the constraints are 
presumed to be effective (for example, see equation \eref{gamma}) --- 
if ineffective constraints occur, they can always be made effective 
for use with this technique; in that sense, the technique is actually 
geometrical.  One might ask whether such modification of ineffective 
constraints can cause problems.  It turns out that if ineffective 
constraints occur, then their presence may modify the gauge fixing 
procedure used in conjunction with the original Dirac method in such a 
way that the counting of degrees of freedom differs from the one 
presented above.  In the next section we discuss a simple example that 
shows the difference between Dirac's original treatment, supplemented 
by gauge fixing, and the quotienting method we've outlined here, which 
corresponds to Dirac's extended method.

Dirac's extended method, which is equivalent to the one we've 
presented here, produces a final phase space which is always even 
dimensional.  Dirac's original procedure, supplemented by gauge 
fixing, has the superiority of being equivalent to the Lagrangian 
variational principle.  Therefore, in spite of the fact that this 
latter method may result in a system with an odd number of degrees of 
freedom (as in the example in the following section), it is to be 
preferred for classical models.
 

\section{Example}
\label{sec.example}

Consider the Lagrangian
\begin{equation}
    L = {1\over2}{\dot x}^2 + {1\over2z}{\dot y}^2\ ,
    \label{lag.example} 
\end{equation} 
where $z\neq 0$.  The Noether gauge 
transformations are 
\[  \delta x=0\ ,\ \delta y= {\epsilon{\dot y}\over z}\ ,\ 
    \delta z={\dot \epsilon}\ ,
\]
where $\epsilon$ is an arbitrary function.

First, we analyze this system from a Lagrangian point of view.  The 
equations of motion are
\begin{equation}
    {\ddot x}=0\ ,\  {\dot y}=0\ .
\end{equation}
The $z$ variable is completely arbitrary and is pure gauge.  These 
equations define a system with three degrees of freedom in tangent 
space, parameterized by $x(0),{\dot x}(0),y(0)$.  Notice that the 
gauge transformation $\delta y$ vanishes on shell, so $y$ is a weakly 
gauge invariant quantity.

Let us now analyze this system using Dirac's method.  The Dirac 
Hamiltonian is
\begin{equation}
    H_D = {1\over2} p_x^2 +{1\over2} z p_y^2 
        - \lambda p_z \ ,
\end{equation} 
where $\lambda$ is the Lagrange multiplier for the primary constraint 
$p_z=0$.  The stabilization of $p_z=0$ gives the ineffective 
constraint $p_y^2=0$, and the algorithm stops here.  The gauge 
generator (Batlle \etal 1989, Pons \etal 1997) is
\begin{equation}
    G = \dot\epsilon p_z + {\epsilon\over2} p_y^2\ , 
\end{equation} 
with $\epsilon$ an arbitrary function of time.  

The gauge fixing procedure, Pons and Shepley (1995), has in general 
two steps.  The first is to fix the dynamics, and the second is to 
eliminate redundant initial conditions.  Here, to fix the dynamics we 
can introduce the general gauge-fixing $z-f(t)=0$ for $f$ arbitrary.  
Stability of this condition under the gauge transformations sets 
$\dot\epsilon = 0$.  Since the coefficient of $\epsilon$ in $G$ is 
ineffective, it does not change the dynamical trajectories, and so the 
gauge fixing is complete.  Notice that this violates the standard 
lore, for we have two first class constraints, $p_z=0$ and $p_y=0$ but 
only one gauge fixing.  This totals three constraints that reduce the 
original six degrees of freedom to three: $x(0),p_x(0),y(0)$, the same 
as in the Lagrangian picture.

Instead, if we apply the method of quotienting out the kernel of the 
presymplectic form, we get as a final reduced phase space 
\[  {\cal P}_f = {M_2 \over{(\{-,p_z\},\{-,p_y\})}} \ ,
\] 
where $M_2$ is the surface in phase space defined by $p_z =0,p_y=0$.  
We have $\bomega_2=\mathbf{d}x\wedge\mathbf{d}p_x$ and 
$H_2={1\over2}p_x^2$.  The dimension of ${\cal P}_f$ is 2.

This result, which is different from that using Dirac's method, 
matches the one obtained with the extended Dirac's Hamiltonian, where 
all final first class constraints (in effective form) are added with 
Lagrange multipliers to the canonical Hamiltonian.  Dirac's conjecture 
was that the original Dirac theory and the extended one were 
equivalent.  We conclude that when Dirac's conjecture holds, the 
method of quotienting out the kernel is equivalent to Dirac's, whereas 
if Dirac's conjecture fails, it is equivalent to the extended Dirac's 
formalism.


\section{Conclusions}
\label{sec.conclusions}

In summary, we have the following.

1) We have obtained a local basis for 
${\cal K} = {\rm Ker}(\bomega_{\rm L})$ in configuration-velocity 
space for any gauge theory.  This is new and allows for trivial 
verifications of the properties of $\cal K$ given in the literature.  
To get these results it has been particularly useful to rely on 
Hamiltonian methods.

2) We have obtained as the final reduced phase space the quotient of 
the final Dirac's constraint surface in canonical formalism by the 
integral surface generated by the final first class constraints in 
effective form.  We find the constraint surface ($M_{f}$ in equation 
\eref{phys.space}) through a projectability requirement on the 
Lagrangian energy (or equivalently, on the Hamiltonian) 
rather than through imposing tangency conditions on 
the Hamiltonian flows.  Let us emphasize this point: We do not talk of 
stabilization of constraints but rather projectability of structures 
which are required to formulate the dynamics in a reduced physical 
phase space.

3) We have compared our results with Dirac's procedure.  An agreement 
exists in all the cases when no ineffective Hamiltonian constraints 
appear in the formalism.  If there are ineffective constraints whose 
effectivization is first class, then our results disagree with 
Dirac's, and it turns out that the quotienting algorithm agrees with 
the extended Dirac formalism.  When there are disagreements, the 
origin is in the structure of the gauge generators.  Sometimes the 
gauge generators contain pieces that are ineffective constraints, and 
they, contrary to the usual case, do not call for any gauge fixing.  
Essentially, the variables that are canonically conjugate to these 
first class ineffective constraints are weakly (on shell) gauge 
invariant.  The quotienting reduction method, as well as Dirac's 
extended formulation, eliminates these variables and yields a phase 
space whose variables are strictly (on and off shell) gauge invariant.  
This is the difference with Dirac's original method, supplemented with 
gauge fixing, which is able to retain the weakly gauge invariant 
quantities.  For this reason we feel that this latter technique is 
superior to the quotienting algorithm in these circumstances --- at 
least for classical models.

4) We have produced a simple example that illustrates the failure of 
Dirac's conjecture in the presence of ineffective constraints.  This 
example also shows that, in Dirac's analysis, it is possible to have 
Hamiltonian formulations with an odd number of physical degrees of 
freedom.  We must remark that in Dirac's approach (supplemented with 
gauge fixing) it is not always true that every first class constraint 
eliminates two degrees of freedom: This does not happen if there are 
first class constraints that appear in the stabilization algorithm in 
ineffective form.

5) It is worth mentioning that other reduction techniques, 
specifically the Faddeev and Jackiw method, may also fail to reproduce 
Dirac's theory (Garc\'\i{}a and Pons 1998) if the formalism contains 
ineffective constraints.

6) Of course, one should not forget quantum mechanics.  The canonical 
approach to quantum mechanics involves a (nonsingular) symplectic 
form, Isham (1984).  In this method, it is therefore required that 
phase space be even-dimensional.  This argument would tend to favor 
the quotienting algorithm.  However, it may be that other approaches 
to quantum mechanics, possibly the path integration approach, do not 
need such a requirement.  And in any case, it is not strictly 
necessary that a model which is acceptable as a classical model be 
quantizable.  It is for these reasons that we say that an approach to 
Hamiltonian dynamics which results in a phase-space picture equivalent 
to the tangent space picture --- the original Dirac method 
supplemented with gauge fixing --- is favored for classical models.


\ack

We are pleased to thank C\'ecile DeWitt-Morette for her advice.  
JMP and DCS would like to thank the Center for Relativity of The 
University of Texas at Austin for its hospitality.  JMP acknowledges 
support by CIRIT and by CICYT contracts AEN95-0590 and GRQ 93-1047 
and wishes to thank the Comissionat per a Universitats i Recerca 
de la Generalitat de Catalunya for a grant.  DCS acknowledges 
support by National Science Foundation Grant PHY94-13063.


\pagebreak

\References

\item[]
Abraham R and Marsden J E 1978
    {\it Foundations of Mechanics} 2nd edn
    (Reading MA: Benjamin-Cummings)

\item[]
Batlle C, Gomis J, Gr\`acia X and Pons J M 1989
    Neother's theorem and gauge transformations:  Application to the 
    bosonic string and $C\!P^{n-1}_{2}$ 
    \JMP {\bf 30} 1345-50

\item[]
Batlle C, Gomis J, Pons J M and Roman N 1986 
    Equivalence between the Lagrangian and Hamiltonian formalism 
    for constrained systems
    \JMP {\bf 27} 2953-62

\item[]
Bergmann P G 1949
    Non-linear field theories
    \PR {\bf 75} 680-5
    
\item[]
Bergmann P G and Goldberg I 1955
   Dirac bracket transformations in phase space
   \PR {\bf 98} 531-8

\item[]
Cantrjin F, Cari\~{n}ena J F, Crampin M and Ibort L A 1986
    Reduction of degenerate Lagrangian systems
    {\it J.\ Geom.\ Phys} {\bf 3} 353-400

\item[]
Cari\~{n}ena J F 1990
    Theory of singular Lagrangians
    {\it Forstsch.\ Phys.} {\bf 38} 641-79
    and references therein

\item[]
Cari\~{n}ena J F, L\'{o}pez C and Rom\'{a}n-Roy N 1988
    Origin of the Lagrangian constraints and their relation with the 
    Hamiltonian formulation
    \JMP {\bf 29} 1143-9

\item[]
Dirac P A M 1950 
    Generalized Hamiltonian dynamics
    {\it Can.\ J.\ Math.} {\bf 2} 129-48

\item[]
\dash 1964
    {\it Lectures on Quantum Mechanics}
    (New York: Yeshiva University Press)

\item[]
Faddeev L and Jackiw R 1993
    Hamiltonian reduction of unconstrained and constrained systems
    \PRL {\bf 60} 1692-4
 
\item[]
Garc\'\i{}a J A and Pons J M 1997 
    Equivalence of Faddeev-Jackiw and Dirac approaches for gauge 
    theories
    {\it Int.\ J.\ Mod.\ Phys.} A {\bf12} 451-64 hep-th/9610067

\item[]
\dash 1998
    Faddeev-Jackiw approach to gauge theories and ineffective 
    constraints
    {\it Int.\ J.\ Mod.\ Phys.} A to be published

\item[]
Gotay M 1982
    On coisotropic imbeddings of presymplectic manifolds
    {\it Proc.\ Am.\ Math.\ Soc.} {\bf 84} 111-4

\item[]
Gotay  M J and Nester J M 1979
    Presymplectic Lagrangian systems I: the constraint algorithm and 
    the equivalence theorem
    {\it Ann.\ Inst.\ H.\ Poincar\'e} A {\bf 30} 129-42

\item[]
\dash 1980 
    Presymplectic Lagrangian systems II: the second-order equation 
    problem
    {\it Ann.\ Inst.\ H.\ Poincar\'e} A {\bf 32} 1-13

\item[]
Gotay M J, Nester J M and Hinds G 1978
    Presymplectic manifolds and the Dirac-Bergmann theory of 
    constraints
    \JMP {\bf 19} 2388-99

\item[]
Gotay  M and Sniatycki J 1981
    On the quantization of presymplectic dynamical systems via 
    coisotropic imbeddings
    {\it Commun.\ Math.\ Phys.} {\bf 82} 377-89

\item[]
Gr\`acia X and Pons J M 1989 
    On an evolution operator connecting Lagrangian and Hamiltonian 
    formalisms
    {\it Lett.\ Math.\ Phys.} {\bf 17} 175-80
	
\item[]
Ibort L A, Landi G, Mar\'{\i}n-Solano J and Marmo G 1993
    On the inverse problem of Lagrangian supermechanics 
    {\it Int.\ J.\ Mod.\ Phys.} A {\bf 8} 3565-76

\item[]
Ibort L A and Mar\'{\i}n-Solano J 1992
    A geometric classification of Lagrangian functions and the 
    reduction of evolution space
    \JPA {\bf 25} 3353-67

\item[]
Isham C J 1984
    Topological and Global Aspects of Quantum Theory
    {\it Relativit\'e, Groupes, et Topologie II} 
    ed DeWitt  B S and Stora R 
    (Amsterdam:  North-Holland) pp~1059-290

\item[]
Jackiw R 1995
    (Constrained) quantization without tears 
    {\it Proc. 2nd Workshop on Constraints Theory and 
    Quantization Methods (Montepulciano, 1993)} 
    (Singapore: World Scientific) pp~163-75 hep-th/9306075

\item[]
Lee J and Wald R M 1990
    Local symmetries and constraints
    \JMP {\bf 31} 725-43

\item[]
Pons J M 1988
    New relations between Hamiltonian and Lagrangian constraints
    \JPA {\bf21} 2705-15

\item[]
Pons J M, Salisbury D C, and Shepley L C 1997
    Gauge transformations in the Lagrangian and Hamiltonian 
    formalisms of generally covariant systems
    \PR D {\bf55} 658-68 gr-qc/9612037.

\item[]
Pons J M and Shepley L C 1995 
    Evolutionary laws, initial conditions and gauge fixing in 
    constrained systems
    \CQG {\bf 12} 1771-90 gr-qc/9508052

\item[]
Sniatycki J 1974
    Dirac brackets in geometric dynamics
    {\it Ann.\ Inst.\ H.\ Poincar\'e} A {\bf 20} 365-72

\endrefs

\end{document}